\documentclass[twocolumn,showpacs,amssymb,pra,nofootinbib]{revtex4}
\usepackage{graphicx}

\usepackage[T1]{fontenc}
\usepackage{mathptmx}

\newcommand{\ket}[1]{| #1 \rangle} 
\newcommand{\bra}[1]{\langle #1 |} 

\begin{document}

\title{Bit Transmission Probability Maximizing the Key Rate of the BB84 Protocol}

\author{Sonny Lumbantoruan}
 \email{sonny@chem-is-try.org}
\author{Ryutaroh Matsumoto}%
 \email{ryutaroh@rmatsumoto.org}
 \homepage{http://www.rmatsumoto.org/research.html}
\author{Tomohiko Uyematsu}
 \email{uyematsu@ieee.org}
 \affiliation{%
Department of Communications and Integrated Systems \\
Tokyo Institute of Technology \\
2-12-1, Oookayama, Meguro-ku, Tokyo, 152-8552, Japan
}

\date{March 11, 2010}

\begin{abstract}
In all papers on the BB84 protocol, the transmission probability of each bit
value is usually set to be equal.
In this paper, we show that by assigning different transmission probability to
each transmitted qubit within a single polarization basis, we can generally
improve the key generation rate of the BB84 protocol and achieve a higher key
rate.
\end{abstract}

\maketitle

\section{Introduction}
Quantum Key Distribution (QKD) has attracted great attention as an
unconditionally secure key distribution scheme.
The basic idea of QKD protocol is to exploit the quantum mechanical principle
that observation in general disturbs the system being observed.
Thus, if there is an eavesdropper (Eve) listening while the two legitimate
communicating users, namely Alice and Bob, attempt to
transmit their key, the presence of the eavesdropper will be visible as a
disturbance of the communication channel that Alice and Bob are
using to generate the secret key.
Alice and Bob can then throw out the key bits established while Eve was
listening in, and start over.
The key generation rate, which is the length of the securely sharable key per
channel use, is one of the most important criteria for the efficiency of
the QKD protocol.
The first QKD protocol, which was proposed in 1984 \cite{BB84}, is called BB84 
after its inventors (Bennet and Brassard).

QKD protocol usually consists of two parts: a quantum and a classical part.
In the quantum part, Alice sends qubits prepared in certain states to Bob.
The states of these qubits are encodings of bit values randomly chosen by Alice.
Bob performs a measurement on the qubits to decode the bit values.
For each of the bits, both the encoding and decoding are chosen from a certain
set of operators. 
After the transmission steps, Alice and Bob apply \textit{sifting} where they
publicly compare the encoding and decoding operator they have used and keep
only the bit pairs for which these operators \textit{match}.

Once Alice and Bob have correlated bit strings, they proceed with the classical
part of the protocol.
In a first step, called \textit{parameter estimation}, they compare the bit
values for randomly chosen samples from their strings to estimate the quantum
channel.
After the parameter estimation, Alice and Bob proceed with a classical
processing, where Alice and Bob share a secret key based on their bit sequences
obtained in the quantum part.

Mathematically, quantum channels are described by trace preserving completely
positive (TPCP) maps \cite{NielsenChuang}.
Conventionally, in the BB84 protocol, we only used the statistics of the matched
measurement outcomes, which are transmitted and received in the same basis, to
estimate the TPCP map that describes the quantum channel,
while the mismatched measurement outcomes, which are transmitted and
received in different bases, were discarded.
However, Watanabe \textit{et al.} \cite{Watanabe} showed that by using
the statistics of \textit{both} matched and mismatched measurement outcomes,
the TPCP maps describing the quantum channel can be estimated more accurately.
They implemented a practical classical processing for the six-state and BB84
protocols that utilizes their accurate channel estimation method and showed that
the key rates obtained with their method were at least as high as the key rates
obtained with the standard processing by Shor and Preskill \cite{ShorPreskill}.

In the BB84 protocol \cite{BB84}, Alice creates random bits of 0 and 1 with
equal probability.
Then, Alice and Bob each chooses between the two bases, i.e.\ the rectilinear
basis (or ${z}$ basis) of vertical ($0^\circ$) and horizontal ($90^\circ$)
polarizations, and the diagonal basis (or ${x}$ basis) of $45^\circ$ and
$135^\circ$ polarizations, with equal probability.
Lo \textit{et al.} \cite{Lo2003} proposed a simple modification of the
standard BB84 protocol \cite{BB84} by assigning significantly different
probabilities to the different polarization bases during both transmission and
reception.
They showed that the modification could reduce the fraction of
mismatched measurement outcomes, thus nearly doubles the efficiency of the
BB84 protocol.

In this paper, we propose a modification of the BB84 protocol by assigning a
different transmission probability to each transmitted qubit \textit{within}
a single polarization basis.
While in classical information, assignment of different probability to each
input bit can increase the mutual information of asymmetric channels
\cite[Problem 7.8]{Cover}, in quantum key distribution the benefit of
assigning a different transmission probability to each transmitted qubit
was unknown.
We show that by setting a different transmission probability to each transmitted
qubit, we can improve the key rate and achieve a higher key rate.
We demonstrate this fact by using the accurate channel estimation over the
amplitude damping channel.
We determine the optimum bit transmission probability that maximizes the key
rate.

\section{Modification of BB84 Protocol}

In this section, we describe a modification of the BB84 protocol
where the transmission probability of each qubit within a single polarization
basis is not necessarily equal.
The protocol consists of a quantum and a classical part.
The quantum part includes the distribution and measurement of quantum
information, and is determined by the operators that Alice and Bob use for
their encoding and decoding.

For simplicity, we assume that Eve's attack is the collective attack\footnote{
By using the de Finetti representation arguments \cite{Ren05, Ren07}, the
result can be extended to the coherent attack.}, i.e.\ the
channel connecting Alice and Bob is given by tensor products of a channel
$\mathcal{E}_B$ from a qubit density matrix to itself.
As is usual in a lot of QKD literature, we assume  that Eve can access all the
environment of channel $\mathcal{E}_B$.
The channel to the environment is denoted by $\mathcal{E}_E$.

\subsection{Quantum part: Distribution of Quantum Information and Measurement}
\label{quantum-part}

In the modified BB84 protocol, Alice chooses random bits of 0 and 1 according to 
the probability distribution
\begin{equation}
P_X(0) := q, \quad \quad P_X(1) := 1-q. \label{probX}
\end{equation}
Alice modulates each bit into a transmission basis that is randomly chosen
from the ${z}$ basis $\{\ket{0_{z}}, \ket{1_{z}}\}$ and the
${x}$ basis $\{\ket{0_{x}}, \ket{1_{x}}\}$, where $\ket{0_{a}}$
and $\ket{1_{a}}$ are the eigenstates of the Pauli matrix
$\sigma_{a}$ for ${a} \in \{{x}, {z}\}$.
We occasionaly omit the subscripts $\{{x}, {y}, {z}\}$
of the basis, and the basis $\{\ket{0}, \ket{1}\}$ is regarded as ${z}$
basis unless otherwise stated.
Then Bob randomly chooses one of the measurement observables $\sigma_{a}$
for ${a} \in \{{x}, {z}\}$,
and converts a measurement result $+1$ or $-1$ into a bit $0$ or $1$,
respectively.
Note that Alice and Bob keep the the mismatched measurement outcomes
to estimate the channel more accurately.

\subsection{Classical part: Parameter Estimation and Classical Processing}
\label{classical-part}

The classical part of the protocol that we consider is essentially the same as
the classical part of the protocol proposed by Watanabe \textit{et al.} \cite{Watanabe}.
However, since we assign a different transmission probability
to each transmitted qubit with a single polarization basis
(see Eq. (\ref{probX})), some adjustments need to be made accordingly.

The classical part of our protocol consists of two subprotocols, called
\textit{parameter estimation} and \textit{classical post-processing}.
The main purpose of the parameter estimation subprotocol is to estimate the
amount of information gained by the eavesdropper Eve during the distribution of
the quantum information.

After the parameter estimation, Alice and Bob proceed with a classical
subprotocol.
Hereafter, we treat only Alice's bit sequence $\vec{x} \in \mathbb{F}_2^n$
that is transmitted in ${z}$ basis and the corresponding Bob's bit
sequence $\vec{y} \in \mathbb{F}_2^n$ that is received in $\sigma_{z}$
measurement, where $\mathbb{F}_2$ is the finite field of order 2.
Our goal is to generate a secure key pair $(\textbf{S}_A, \textbf{S}_B)$,
using $\vec{x}$ and $\vec{y}$.
Here Alice and Bob want to generate a key pair $(\textbf{S}_A, \textbf{S}_B)$
which is statistically independent of Eve's information by cloning the quantum objects
and looking at the conversation over the public authenticated channel.
The protocol we consider is \textit{one-way}, i.e.\ only communication from
Alice to Bob or from Bob to Alice, is needed.
It consists of the following steps:
\begin{enumerate}
\item \textit{Information Reconciliation}: Alice sends error correction
information to Bob. Using the correction information, Bob decode the bit
string $\vec{y}$ into an estimate of $\vec{x}$.
\item \textit{Privacy Amplification}: Alice randomly chooses a hash function
from a set of universal hash functions and sends the choice to Bob over the
public channel. Then, Alice and Bob compute $\textbf{S}_A$ and $\textbf{S}_B$,
respectively.
\end{enumerate}

The above procedure is usually called \textit{the direct reconciliation}.
The procedure in which the roles of Alice and Bob are switched is called
\textit{the reverse reconciliation} \cite{19Maurer}.

Since the pair of the sequences $(\vec{x}, \vec{y})$ is transmitted and
received in ${z}$ basis, they are independently identically distributed
according to
\begin{equation}
P_{XY}(x,y) := P_X(x) \bra{y_{z}} \mathcal{E}_B(\ket{x_{z}}\bra{x_{z}}) \ket{y_{z}}. \label{jointpr}
\end{equation}
Note that the distribution $P_{XY}$ can be estimated from the statistics of the
sample bits that are transmitted in ${z}$ basis and measured by the
observable $\sigma_{z}$.

The secure key rate is determined according to the result of the privacy
amplification \cite{Ren05}.
For the direct reconciliation, let
\begin{equation}
H_\rho(X|E) := H(\rho_{XE}) - H(\rho_E)
\end{equation}
be the conditional von Neumann entropy with respect to density matrix
\begin{equation}
\rho_{XE} := \sum_{x \in \mathbb{F}_2} P_X(x) \ket{x}\bra{x}\otimes\mathcal{E}_E(\ket{x}\bra{x}),
\end{equation}
where $H(\rho)$ is the von Neumann entropy for a density matrix $\rho$ and
$P_X(x)$ is the probability distribution shown in Eq. (\ref{probX}).
The secure key rate \cite{Ren05} is
\begin{equation}
H_\rho(X|E) - H(X|Y).
\end{equation}

While for the reverse reconciliation, we can calculate the conditional
von Neumann entropy $H_\rho(Y|E):=H(\rho_{YE})-H(\rho_E)$ from
the channel $\mathcal{E}_B$ as follows.
We define
\begin{equation}
\rho_{AB} := (I \otimes \mathcal{E}_B)(\ket{\psi}\bra{\psi}), \label{rhoAB-def}
\end{equation}
for the entangled state
\begin{eqnarray}
\ket{\psi} := \sqrt{q} \ket{00} + \sqrt{1-q} \ket{11}.
\end{eqnarray}
Let $\Psi_{ABE}$ be a purification of
$\rho_{AB}$, and let 
$\rho_{BE} := \mathrm{tr}_A \left[\Psi_{ABE}\right]$.
Then the density matrix $\rho_{YE}$
is derived by measurement on Bob's system, i.e.
\begin{equation}
\rho_{YE} := \sum_{x \in \mathbb{F}_2} (\ket{y}\bra{y}\otimes I)\rho_{BE}(\ket{y}\bra{y}\otimes I).
\end{equation}
For the reverse reconciliation, the secure key rate \cite{Ren05} is
\begin{equation}
H_\rho(Y|E) - H(Y|X).
\end{equation}

\section{Evaluation of Key Rate}

\subsection{Estimation of Eve's Ambiguity}

In the Stokes parametrization, the qubit channel $\mathcal{E}_B$ can be
described by the affine map parametrized by 12 real parameters \cite{38,39}
as follows:
\begin{equation}
\left[
\begin{array}{c}
\theta_{z}  \\
\theta_{x}  \\
\theta_{y} 
\end{array}
\right]\mapsto
\left[
\begin{array}{ccc}
R_{zz} & R_{zx}& R_{zy} \\
R_{xz} & R_{xx}& R_{xy} \\
R_{yz} & R_{yx}& R_{yy}
\end{array}
\right]
\left[
\begin{array}{c}
\theta_{z}  \\
\theta_{x}  \\
\theta_{y} 
\end{array}
\right]+
\left[
\begin{array}{c}
t_{z}  \\
t_{x}  \\
t_{y} 
\end{array}
\right], \label{channelparam}
\end{equation}
where $(\theta_{z},\theta_{x},\theta_{y})$ describes a vector
in the Bloch sphere \cite{NielsenChuang}.
When Alice and Bob use only ${z}$ basis and ${x}$ basis, the
statistics of the input and output are irrelevant to the parameters
$(R_{zy}, R_{xy}, R_{yz}, R_{yx}, R_{yy}, t_{y})$ in Eq. (\ref{channelparam}).
Thus we can only estimate the parameters
$\omega = (R_{zz}, R_{zx}, R_{xz}, R_{xx}, t_{z}, t_{x})$ by the accurate channel estimation \cite{Watanabe}
and we have to consider the worst case \cite{Watanabe} for the parameters $\omega$,
i.e.
\begin{equation}
F(\omega) := \min_{\tau \in \mathcal{P}'(\omega)} H_{\rho_\tau}(X|E), \label{fomega1}
\end{equation}
where $\mathcal{P}'(\omega)$ is the set of all parameters
$\tau = (R_{zy}, R_{xy}, R_{yz}, R_{yx}, R_{yy}, t_{y})$
such that the parameters $\omega$ and $\tau$ constitute a qubit channel,
and $\rho_\tau$ is the density matrix which corresponds to the parameter $\tau$.

We can simplify the form of the desired function $F(\omega)$
when Eve's ambiguity is convex \cite{Watanabe}.
We can prove the convexity of Eve's ambiguity with
respect to $\mathcal{E}_B$ in our protocol
by using the same technique used by Watanabe \textit{et al.} \cite[Lemma 2]{Watanabe}.
By the convexity of Eve's ambiguity, the minimization in Eq.\ (\ref{fomega1})
is achieved when the parameters
$R_{{zy}}$, $R_{{xy}}$, $R_{{yz}}$, $R_{{yx}}$,
and $t_{{y}}$, are all $0$ \cite[Proposition 1]{Watanabe}. Hence, the number of
free parameters can be reduced to 1 and the remaining free parameter is
$R_{yy}$.
Thus the the problem is rewritten as looking for an estimator of
\begin{equation}
F(\omega) = \min_{R_{yy} \in \mathcal{P}(\omega)} H_{\rho_{R_{yy}}} (X|E),
\end{equation}
where $\mathcal{P}(\omega)$ is the set of parameters $R_{yy}$
such that the parameters $\omega$ and $R_{yy}$ consitute a qubit
channel when other parameters are all $0$, and $\rho_{R_{yy}}$ is
the density matrix corresponding to the parameter $R_{yy}$.

\subsection{Key Rates of Amplitude Damping Channel}

In this section, we calculate the key rates of the BB84 protocol
with our proposed procedure over the amplitude damping channel, and determine
the optimum bit transmission probability that maximizes the key generation rate.
We clarify the fact that the key rates using the optimum bit transmission
probability of the proposed BB84 protocol is higher than those of the
conventional protocol \cite{Watanabe}.

In the Stokes parametrization, the amplitude damping channel $\mathcal{E}_p$
is given by the affine map
\begin{equation}
\label{eq:amplitude-damping}
\left[
\begin{array}{c}
\theta_{z}  \\
\theta_{x}  \\
\theta_{y} 
\end{array}
\right]\mapsto
\left[\begin{array}{ccc}
\hspace{-1mm}1-p\hspace{-1mm}&\hspace{-1mm}0\hspace{-1mm}&\hspace{-1mm}0\hspace{-1mm}\\
\hspace{-1mm}0\hspace{-1mm}&\hspace{-1mm}\sqrt{1-p}\hspace{-1mm}&\hspace{-1mm}0\hspace{-1mm}\\
\hspace{-1mm}0\hspace{-1mm}&\hspace{-1mm}0\hspace{-1mm}&\hspace{-1mm}\sqrt{1-p}\hspace{-1mm}%
\end{array}\right]
\left[
\begin{array}{c}
\theta_{z}  \\
\theta_{x}  \\
\theta_{y} 
\end{array}
\right]+
\left[
\begin{array}{c}
p  \\
0  \\
0 
\end{array}
\right],
\end{equation}
where $0 \le p \le 1$.

In the BB84 protocol, we can estimate the parameters
$R_{zy} = 1-p$, $R_{zx} = 0$, $R_{xz} = 0$, $R_{xx} = \sqrt{1-p}$, $t_{z} = p$, and $t_{x} = 0$.
As explained in the previous section, we can set $R_{zy} = R_{xy} = R_{yz} = R_{zy} = R_{yx} = t_{y} = 0$.
Furthermore, by the condition on the TPCP map \cite{39}
\begin{equation}
(R_{xx} - R_{yy})^2 \leq (1 - R_{zz})^2 - t_{z}^2,
\end{equation}
we can decide the remaining parameter as $R_{yy} = \sqrt{1-p}$. 


By straightforward calculation, the asymptotic key generation rates for the
direct and reverse reconciliations are
\begin{eqnarray}
h\Big(q + p(1-q)\Big) - h\Big(p(1-q)\Big)\label{ratedirect}
\end{eqnarray}
and
\begin{eqnarray}
h(q) - h\Big(p(1-q)\Big),\label{ratereverse}
\end{eqnarray}
respectively, where $h(\bullet)$ is the binary entropy function.


From Eqs. (\ref{ratedirect}) and (\ref{ratereverse}), we can easily see for $p = 0$,
the asymptotic key generation rates for both direct and reverse reconciliations
reach the maximum value when $q = \frac{1}{2}$.
Please recall that $q$ is the bit transmission probability of bit 0 (see Eq. (\ref{probX})). 

We can derive the optimum bit transmission probability by the extreme value theorem.
Let $\hat{q}$ be the optimum bit transmission probability,
i.e.\ the bit transmission probability (of bit 0) that maximizes the
key generation rate such that the key generation rate is positive.
Then the channel parameter $p$
and the optimum bit transmission probability $\hat{q}$ satisfy the following
condition:
\begin{itemize}
\item For direct reconciliation
\begin{equation}
\frac{1 - \hat{q}}{\hat{q}} = \left(\frac{p(1 - \hat{q})}{1 - p(1 - \hat{q})}\right)^p,\label{relationpq-reverse}
\end{equation}
where $0\leq p < 1$ and $0 < \hat{q} < 1$.

\item For reverse reconciliation
\begin{equation}
\frac{1 - p(1 - \hat{q})}{p(1 - \hat{q})} = \left(\frac{q + p(1 - \hat{q})}{(1 - p)(1 - \hat{q})}\right)^\frac{1-p}{p},\label{relationpq-direct}
\end{equation}
where $0\leq p < \frac{1}{2}$ and $0 < \hat{q} < 1$.
\end{itemize}

\begin{figure}
\includegraphics[width=\linewidth]{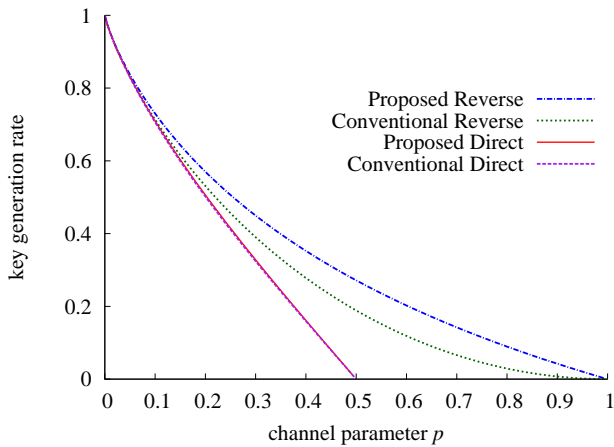}
\caption{(Color online) Comparison of the asymptotic key generation rates against the
channel parameter $p$ of the amplitude damping channel $\mathcal{E}_p$.
``Proposed Reverse" and ``Proposed Direct" are the maximum asymptotic key
generation rates for the reverse and direct reconciliations with the optimum bit
transmission probability $\hat{q}$, respectively.
While ``Conventional Reverse" and
``Conventional Direct" are the asymptotic key generation rates for the reverse and direct reconciliations when $q = \frac{1}{2}$, respectively given in \cite{Watanabe}.}
\label{comparison}
\end{figure}

The key rates for the direct and reverse reconciliations using the optimum bit
transmission probability are plotted in Fig. \ref{comparison}.
We find that the proposed key rates, i.e.\ the key rates when $q = \hat{q}$,
are higher than the conventional ones \cite{Watanabe}, i.e.\ the key rates when $q=\frac{1}{2}$,
in both the direct and reverse reconciliations. In the direct reconciliation,
the proposed key rate is slightly higher than that of the conventional one so
that the lines of the two key rates seem to overlap one another.
While in contrast, in the reverse reconciliation, the proposed key rate grows
much higher than the conventional one as the parameter $p$ increases.
And especially when the parameter $p \gtrsim 0.7$, we can see that the
proposed key rate is more than twice as high as the conventional one.

\section{Conclusion}

In this paper, we proposed a simple modification of the BB84 protocol
where the transmission probability of each qubit within a single polarization
basis is not necessarily equal.
We showed that by assigning a different transmission probability to each
transmitted qubit, we can generally increase the key generation rate of the BB84
protocol.
We demonstrated this by using the accurate channel estimation over the amplitude
damping channel.
We determined the optimum bit transmission probability that maximizes the key
generation rate.
We showed that in general, assignment of an equal probability to each qubit
within a single polarization basis is not necessarily optimal in 
QKD protocol.

\section*{Acknowlegdment}
We would like to thank Dr.\ Shun Watanabe for valuable discussions.


\begin{thebibliography}{48}
\expandafter\ifx\csname natexlab\endcsname\relax\def\natexlab#1{#1}\fi
\expandafter\ifx\csname bibnamefont\endcsname\relax
  \def\bibnamefont#1{#1}\fi
\expandafter\ifx\csname bibfnamefont\endcsname\relax
  \def\bibfnamefont#1{#1}\fi
\expandafter\ifx\csname citenamefont\endcsname\relax
  \def\citenamefont#1{#1}\fi
\expandafter\ifx\csname url\endcsname\relax
  \def\url#1{\texttt{#1}}\fi
\expandafter\ifx\csname urlprefix\endcsname\relax\def\urlprefix{URL }\fi
\providecommand{\bibinfo}[2]{#2}
\providecommand{\eprint}[2][]{\url{#2}}

\bibitem[{\citenamefont{Bennett and Brassard}(1984)}]{BB84}
\bibinfo{author}{\bibfnamefont{C.~H.} \bibnamefont{Bennett}} \bibnamefont{and}
  \bibinfo{author}{\bibfnamefont{G.}~\bibnamefont{Brassard}}, in
  \emph{\bibinfo{booktitle}{Proc. IEEE Int. Conf. Computers Systems and Signal
  Processing}} (\bibinfo{address}{Bangalore, India}, \bibinfo{year}{1984}), pp.
  \bibinfo{pages}{175--179}.

\bibitem[{\citenamefont{Nielsen and Chuang}(2000)}]{NielsenChuang}
\bibinfo{author}{\bibfnamefont{M.~A.} \bibnamefont{Nielsen}} \bibnamefont{and}
  \bibinfo{author}{\bibfnamefont{I.~L.} \bibnamefont{Chuang}},
  \emph{\bibinfo{title}{Quantum Computation and Quantum Information}}
  (\bibinfo{publisher}{Cambridge University Press}, \bibinfo{year}{2000}).

\bibitem[{\citenamefont{Watanabe}(1999)}]{Watanabe}
\bibinfo{author}{\bibfnamefont{S.}~\bibnamefont{Watanabe}},
  \bibinfo{author}{\bibfnamefont{R.}~\bibnamefont{Matsumoto}} \bibnamefont{and}
  \bibinfo{author}{\bibfnamefont{T.}~\bibnamefont{Uyematsu}},
  \bibinfo{journal}{Phys. Rev. A} \textbf{\bibinfo{volume}{78}},
  \bibinfo{pages}{042316} (\bibinfo{year}{2008}).

\bibitem[{\citenamefont{Shor and Preskill}(2000)}]{ShorPreskill}
\bibinfo{author}{\bibfnamefont{P.~W.} \bibnamefont{Shor}} \bibnamefont{and}
  \bibinfo{author}{\bibfnamefont{J.}~\bibnamefont{Preskill}},
  \bibinfo{journal}{Phys. Rev. Lett.} \textbf{\bibinfo{volume}{85}},
  \bibinfo{pages}{441} (\bibinfo{year}{2000}),
  arXiv:quant-ph/0003004.

\bibitem[{\citenamefont{Lo, Chau and Ardehali}(2005)}]{Lo2003}
\bibinfo{author}{\bibfnamefont{H.~K.} \bibnamefont{Lo}}, {\bibfnamefont{H.~F.} \bibnamefont{Chau}} \bibnamefont{and} {\bibfnamefont{M.} \bibnamefont{Ardehali}},
  \bibinfo{journal}{J. Cryptol.} \textbf{\bibinfo{volume}{18}},
  \bibinfo{pages}{133--165} (\bibinfo{year}{2005}).

\bibitem[{\citenamefont{Cover and Thomas}(2006)}]{Cover}
\bibinfo{author}{\bibfnamefont{T.~M.} \bibnamefont{Cover}} \bibnamefont{and}
  \bibinfo{author}{\bibfnamefont{J.~A.} \bibnamefont{Thomas}},
  \emph{\bibinfo{title}{Elements of Information Theory}}
  (\bibinfo{publisher}{John Wiley \& Sons}, \bibinfo{year}{2006}),
  \bibinfo{edition}{2nd} ed.

\bibitem[{\citenamefont{Renner et~al.}(2005)\citenamefont{Renner, Gisin, and
  Kraus}}]{Ren05}
\bibinfo{author}{\bibfnamefont{R.}~\bibnamefont{Renner}},
  \bibinfo{author}{\bibfnamefont{N.}~\bibnamefont{Gisin}} \bibnamefont{and}
  \bibinfo{author}{\bibfnamefont{B.}~\bibnamefont{Kraus}},
  \bibinfo{journal}{Phys. Rev. A} \textbf{\bibinfo{volume}{72}},
  \bibinfo{pages}{012332} (\bibinfo{year}{2005}),
  arXiv:quant-ph/0502064.

\bibitem[{\citenamefont{Renner}(2007)}]{Ren07}
\bibinfo{author}{\bibfnamefont{R.}~\bibnamefont{Renner}},
  \bibinfo{journal}{Nature Physics} \textbf{\bibinfo{volume}{3}},
  \bibinfo{pages}{645} (\bibinfo{year}{2007}),
  arXiv:quant-ph/0703069.

\bibitem[{\citenamefont{Maurer}(1993)}]{19Maurer}
\bibinfo{author}{\bibfnamefont{U.}~\bibnamefont{Maurer}},
  \bibinfo{journal}{IEEE Trans. Inform. Theory} \textbf{\bibinfo{volume}{39}},
  \bibinfo{pages}{733} (\bibinfo{year}{1993}).

\bibitem[{\citenamefont{Fujiwara and Nagaoka}(1998)}]{38}
\bibinfo{author}{\bibfnamefont{A.} \bibnamefont{Fujiwara}} \bibnamefont{and} {\bibfnamefont{H.} \bibnamefont{Nagaoka}},
  \bibinfo{journal}{IEEE Trans. Inf. Theory} \textbf{\bibinfo{volume}{44}},
  \bibinfo{pages}{1071} (\bibinfo{year}{1998}).

\bibitem[{\citenamefont{Fujiwara and Algoet}(1999)}]{39}
\bibinfo{author}{\bibfnamefont{A.} \bibnamefont{Fujiwara}} \bibnamefont{and} {\bibfnamefont{P.} \bibnamefont{Algoet}},
  \bibinfo{journal}{Phys. Rev. A} \textbf{\bibinfo{volume}{59}},
  \bibinfo{pages}{3290} (\bibinfo{year}{1999}).

\end{thebibliography}
\end{document}